\documentclass{ws-ijmpe}

\def\eV{\hbox{ eV}}

\def\GeV{\hbox{ GeV}}
\def\TeV{\hbox{ TeV}}
\def\mm{\hbox{ mm}}

\def\y{\hbox{ y}}

\begin{document}

\markboth{M. G\'o\'zd\'z and W. A. Kami\'nski}{Neutrinoless double beta decay
  constrained by the existence of large extra dimensions}

%
\catchline{}{}{}{}{}
%

\title{NEUTRINOLESS DOUBLE BETA DECAY \\
       CONSTRAINED BY THE EXISTENCE \\ OF LARGE EXTRA DIMENSIONS}

\author{MAREK G\'O\'ZD\'Z and WIES{\L}AW A. KAMI\'NSKI}

\address{
Department of Theoretical Physics, Maria Curie-Sk{\l}odowska University, \\ 
Radziszewskigo 10, PL-20-031 Lublin, Poland \\
mgozdz@kft.umcs.lublin.pl \\ 
kaminski@neuron.umcs.lublin.pl}

\maketitle

\begin{history}                %
\received{September 24, 2003}	%
\end{history}               	%

\begin{abstract}
We present the possible influence on the half-life of neutrinoless
double beta decay coming from the existence of $n$ extra spatial
dimensions. The half-life in question depends on the mass of the electron
neutrino. We base our analysis on the Majorana neutrino mass mechanism in
Arkani-Hamed--Dimopoulos--Dvali model.
\end{abstract}

\section{Introduction}

Theories and models with additional spatial dimensions have drawn much
attention during the last few years (see e.g. \cite{kubyshin} and
references therein for a complete review). The idea, which dates back to
the 20's of the previous century, has been recently rediscovered due to
the development of string theory, which requires for a consistent
formulation ten, rather than three, spatial dimensions. Moreover, new
multidimensional generalizations of strings, called branes, emerge from
this theory in a natural way. Recent models suggest that our observable
universe could be embedded in such a brane, which in turn floats in a
higher dimensional bulk, possibly interacting with fields that populate
the bulk as well as with other branes.

The primary goal of the ADD model (\cite{add1,add2,add3} and references
therein) was to explain the huge difference between the scales of
electroweak interactions ($\sim 1 \TeV$) and gravity (i.e. Planck energy
$\sim 10^{16} \TeV$). It is achieved by assuming that the Standard Model
(SM) is localized on a three-dimensional brane which is embedded into a
$(4+n)$-dimensional space-time. The only boson that feels the additional
space (the bulk) is the graviton, and therefore the only interaction
which may freely propagate through the bulk is gravity. This mechanism
gives a natural suppression of that interaction, coming from the volume
of the bulk.

In the present paper we discuss the implications of possible existence
of extra dimensions on an exotic nuclear process: the neutrinoless
double beta decay.
 
\section{Theoretical Background}

Let us assume, that space-time is $(4+n)$-dimensional. Since we do not
observe such a situation in everyday life, the additional $n$ spatial
dimensions must be compactified. For simplicity we will treat them as
curled into circles with very small, identical radii $R$. The whole SM
is restricted to a three-dimensional brane, a topological object, which
is a higher dimensional generalization of a 1-D string and 2-D membrane.

In the ADD model there exists a second brane, parallel to our SM brane,
from which particles carying non-zero lepton number (call them $\chi$)
may escape into the bulk. These particles, in turn, interact with SM
fields on our brane, which leads to a Majorana neutrino mass term,
naturally suppressed by the distance between the branes. We end up with
a mass term of the form \cite{mg-art3}
\begin{equation}
  m_{Maj} \sim v^2 R^\frac{n(n-1)}{n+2} M_{Pl}^\frac{2(1-n)}{n+2}
  \Delta_n(R,m_\chi),
\label{mmaj}
\end{equation}
where $v$ is the vaccum expectation value (vev) of the Higgs field
($v=174\GeV$), and $M_{Pl}$ is the Planck mass. The $n$-dimensional
propagator for the messenger particle with mass $m_\chi$, travelling
between branes, is explicitly
given by
\begin{eqnarray}
  Rm_\chi\ll 1:& \Delta_2(R,m_\chi)    \sim -\log(Rm_\chi), 
  \qquad &       \Delta_{n>2}(R,m_\chi)\sim \frac{1}{R^{n-2}}, 
  \nonumber \\
  Rm_\chi\gg 1:& \Delta_2(R,m_\chi) \sim \frac{e^{-Rm_\chi}}{\sqrt{Rm_\chi}},
  \qquad &       \Delta_{n>2}(R,m_\chi)\sim \frac{e^{-Rm_\chi}}{R^{n-2}}.
\label{delta}
\end{eqnarray}
A small but non-zero Majorana neutrino mass term gives rise to
suppressed lepton number violating processes, like the neutrinoless
double beta decay ($0\nu 2\beta$)
\begin{displaymath}
  A(Z,N) \to A(Z+2,N-2) + 2e^-.
\end{displaymath}
This decay requires that two neutrinos emited in beta decays annihilate
with each other. It is readily seen that this process violates lepton
number by two units, thus is forbidden in the framework of SM. As a
matter of fact, $0\nu 2\beta$ has not been observed, but restrictions on
the amplitude of the decay, which follow from its non-observability, set
valuable constraints on the shape of non-standard physics.

One expects that the half-life of $0\nu 2\beta$ depends on the mass of
the electron neutrino, and that is indeed the case. Under the following
assumptions about the neutrino mass eigenstates: ({\it i}) the
contribution coming from $m_3$ is neglected (this is justified by the
CHOOZ results \cite{chooz}), ({\it ii}) the remaining masses are nearly
degenerate $m_1 \approx m_2$, one obtains for the half-life of the
$0\nu2\beta$ decay \cite{mg-art3}
\begin{equation}
  T_{1/2}^{th} > \kappa \cdot 10^{\frac{93n-150}{n+2}}
  R^{\frac{2n(1-n)}{n+2}} [\Delta_n(R)]^{-2} \y.
\label{t12th}
\end{equation}
Here, the uncertainty factor $\kappa$ satisfies $0.74 < \kappa < 1.17$.
In the derivation of Eq. (\ref{t12th}) the experimental values
established by the IGEX collaboration \cite{igex} $T_{1/2}^{IGEX} > 1.57
\cdot 10^{25}$~y and the effective neutrino mass $\langle m_\nu\rangle =
0.4 \eV$, as suggested by the Heidelberg--Moscow project
\cite{0nu2beta}, have been used.

\section{Results and Discussion}

\begin{figure}
\hbox{
{\includegraphics[width = 0.5\textwidth]{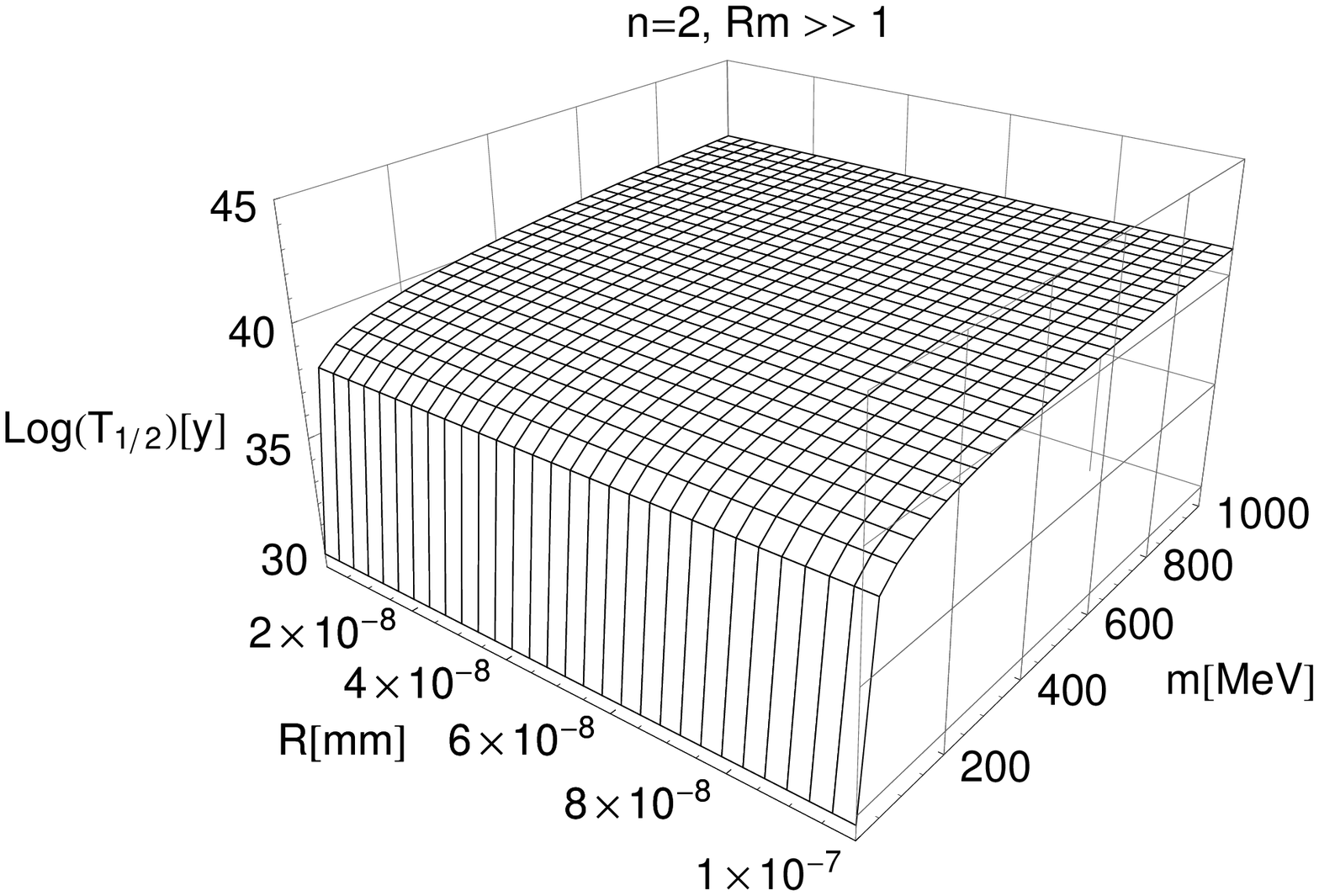}}
{\includegraphics[width = 0.5\textwidth]{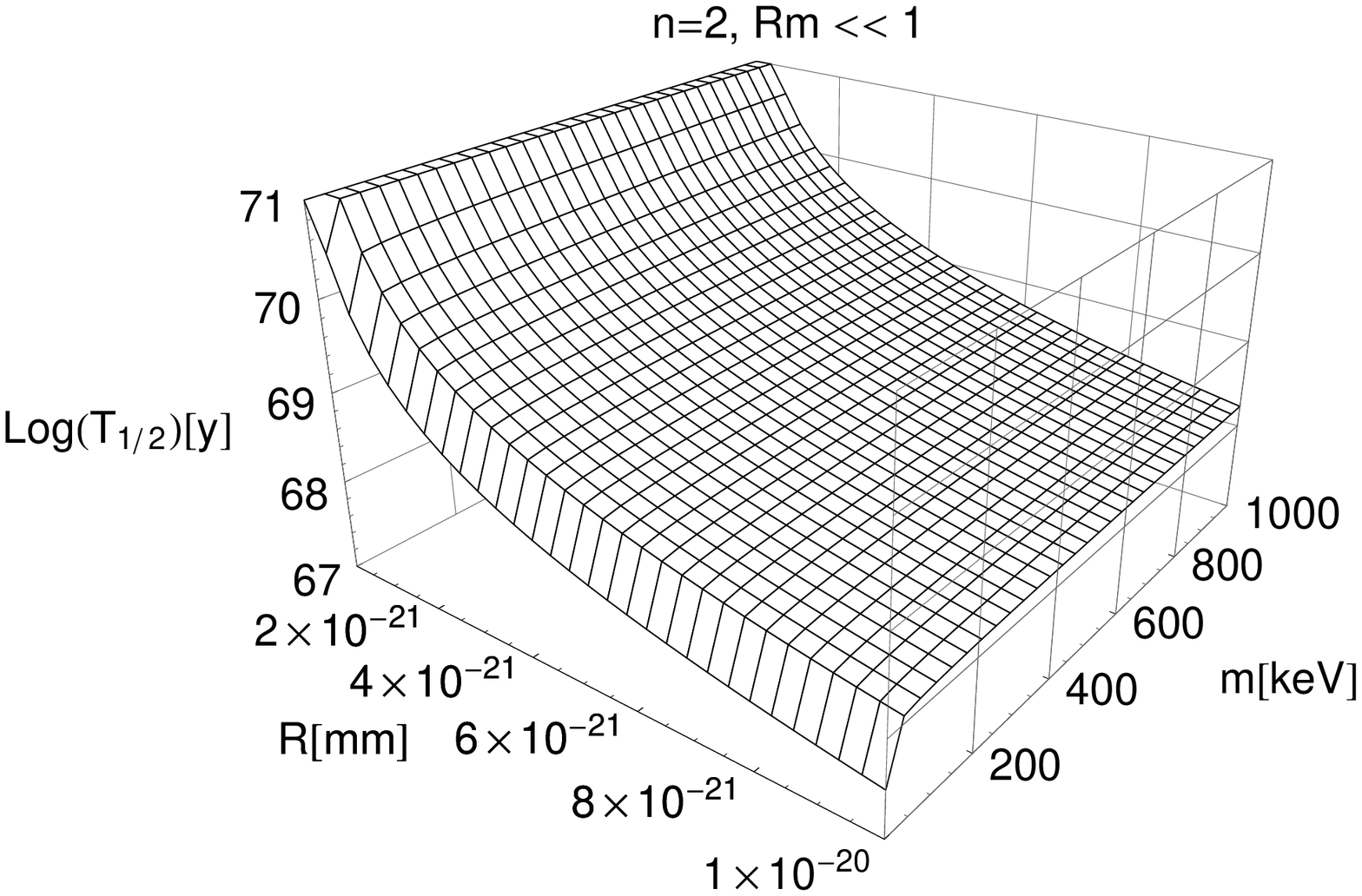}}
}
\caption{\label{fig1} Lower limits on $T_{1/2}$ in the case $n=2$.}
\end{figure}
\begin{figure}
\hbox{
{\includegraphics[width = 0.5\textwidth]{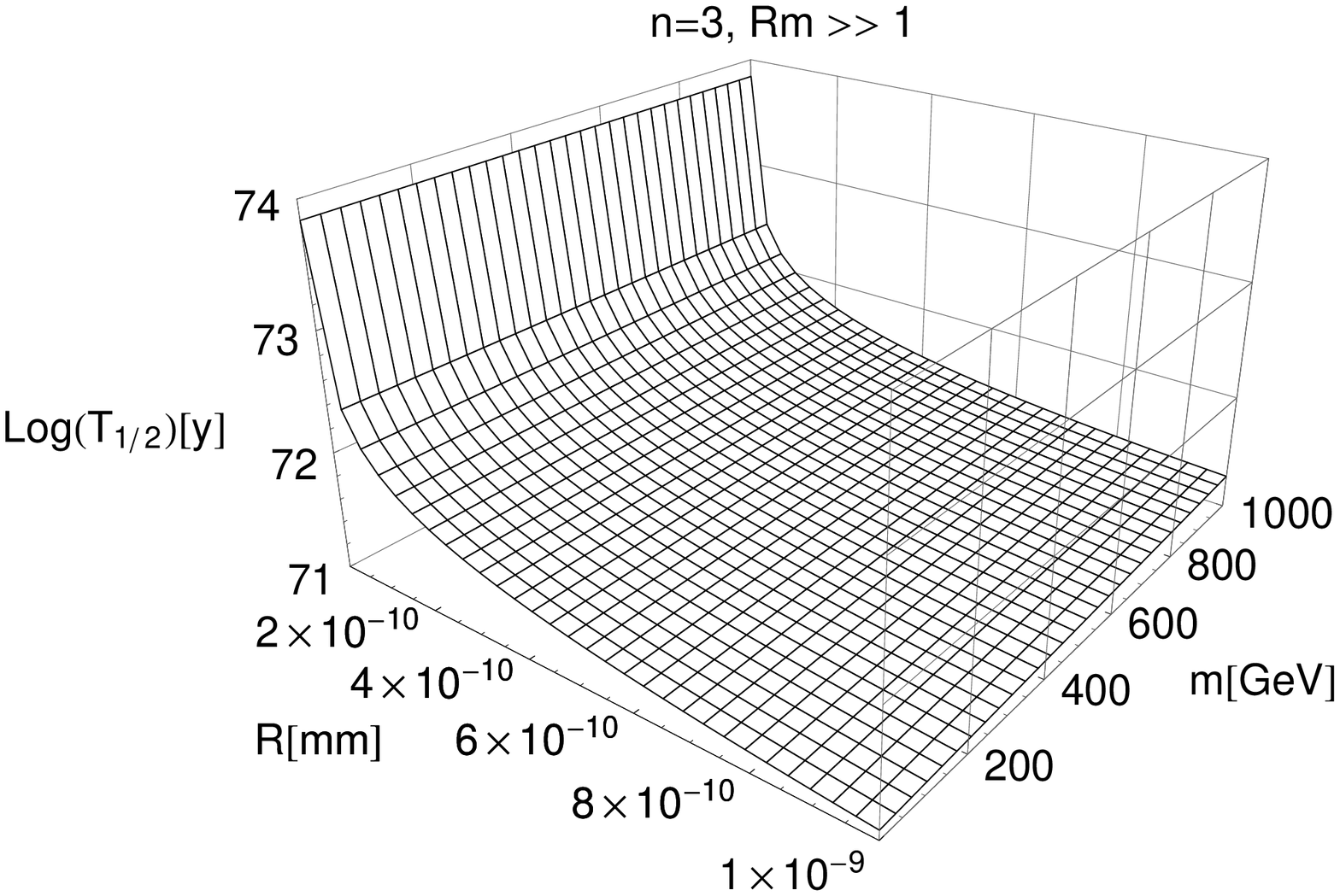}}
{\includegraphics[width = 0.5\textwidth]{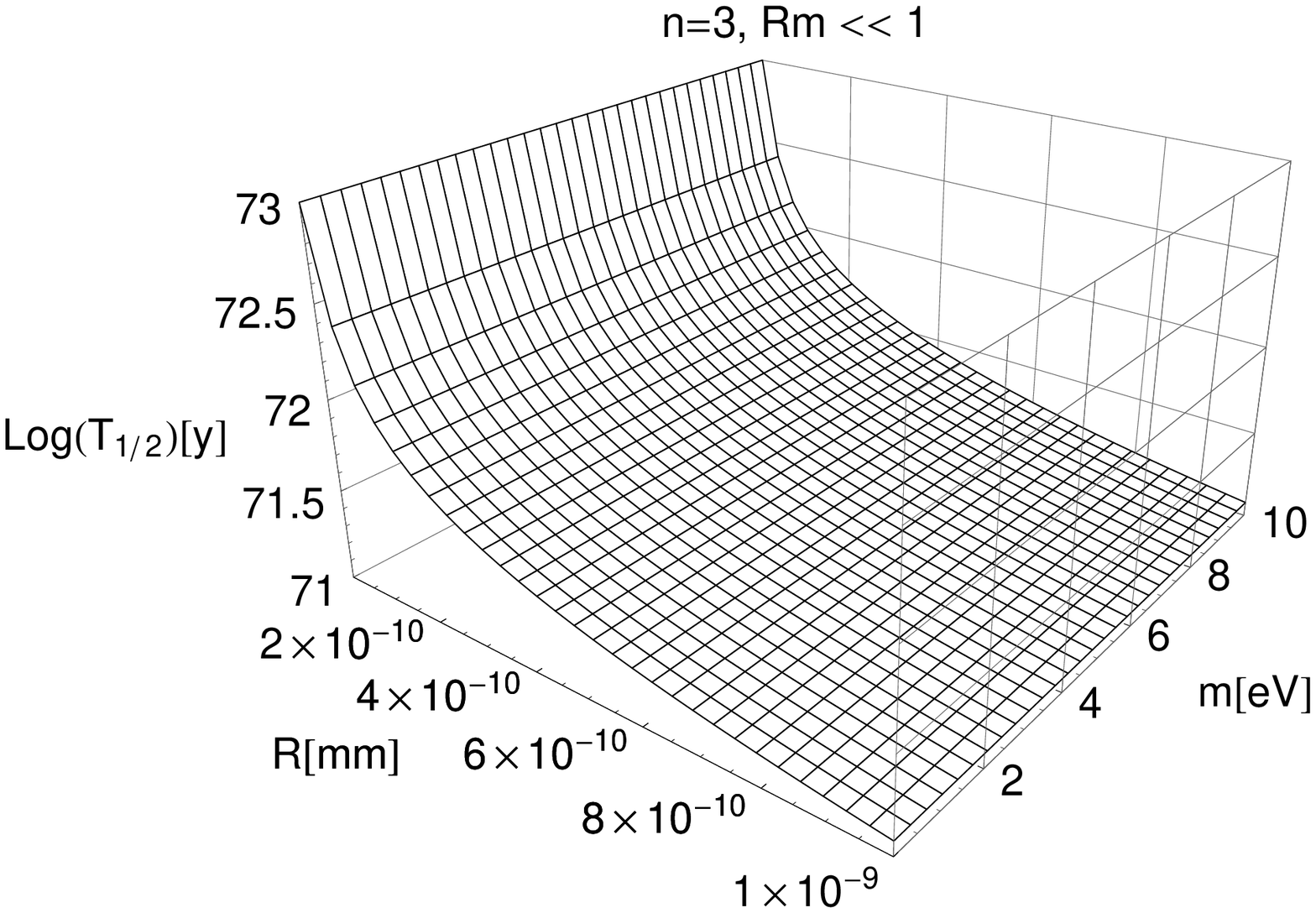}}
}
\caption{\label{fig2} Lower limits on $T_{1/2}$ in the case $n=3$.}
\end{figure}
\begin{figure}
\hbox{
{\includegraphics[width = 0.5\textwidth]{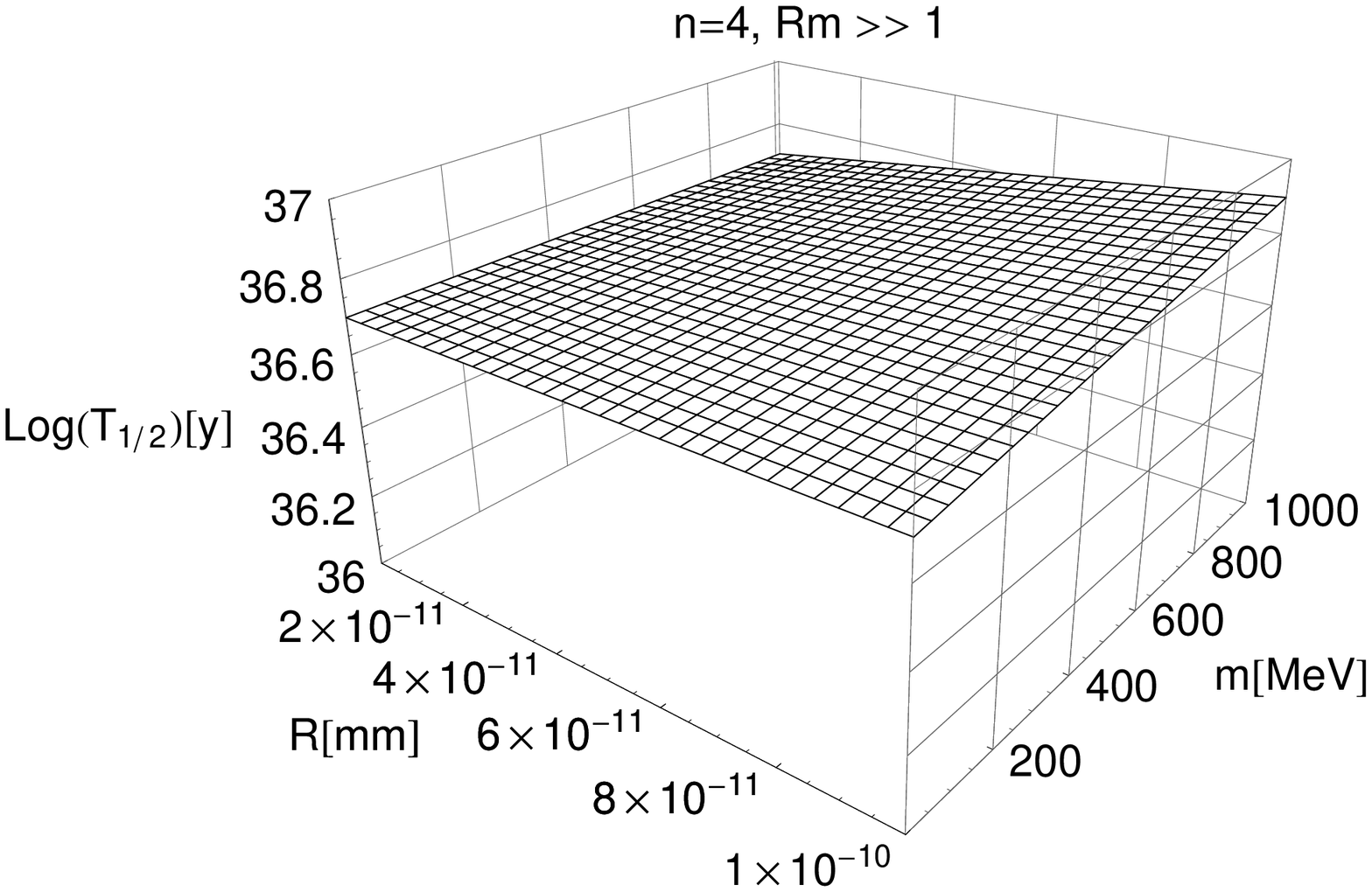}}
{\includegraphics[width = 0.5\textwidth]{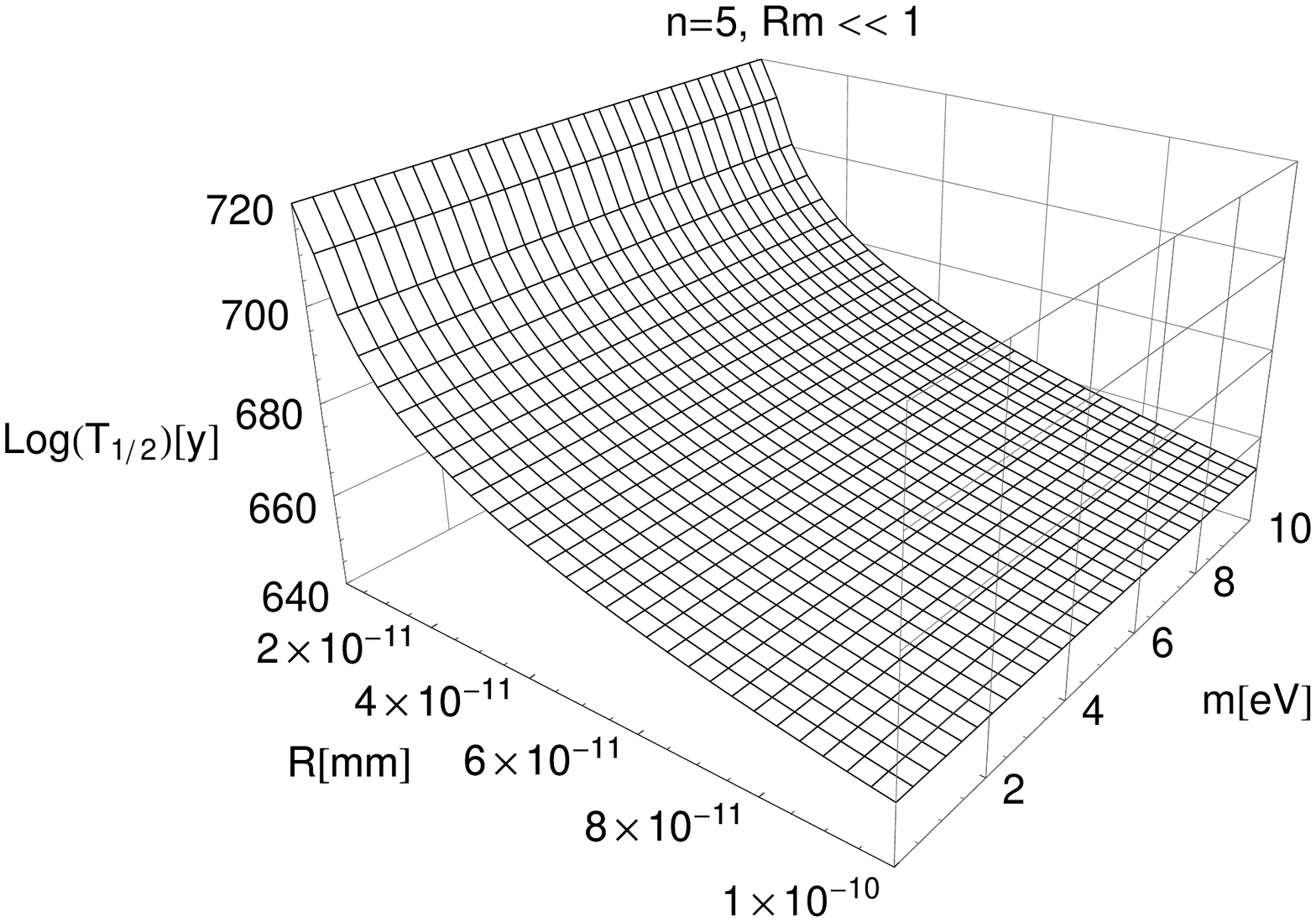}}
}
\caption{\label{fig3} Lower limits on $T_{1/2}$ in the cases $n=4$ and
  $n=5$ (see text for details).}
\end{figure}

We proceed with a detailed analysis of Eq. (\ref{t12th}), taking into
account different cases, according to Eq. (\ref{delta}). One must,
however, be aware of the bounds on $R$ coming from cosmology and
astrophysics. At present they are \cite{hr3} $R < 1.5 \times 10^{-7}\mm$
for $n=2$, $R < 2.6 \times 10^{-9}\mm$ for $n=3$, and $R < 3.4 \times
10^{-10}\mm$ for $n=4$. The same goes to the experimental limit on
$T_{1/2}$ which is at present $T_{1/2}^{IGEX} > 1.57 \times 10^{25}$~y. 

Our results are presented in Figs. \ref{fig1} -- \ref{fig3}. Notice
first, that the dependence on the mass of the messenger is in general
rather weak, except for a narrow region of very small $m_\chi$. We see
that the closest to experimental lower bound is the case of $n=2$ with a
heavy messenger mass $m_\chi$ of a few MeV. Another possibility is $n=4$
and a heavy messenger with a pretty arbitrary mass. The remainig cases
shown in Figs. \ref{fig1} and \ref{fig2} are completely non-verifiable
experimentally. There is also an exceptional case $n=4$ with a light
messenger (not ilustrated in the figures), in which the dependence on
$R$ in Eq. (\ref{t12th}) is lost. Explicitly we get $T_{1/2}\sim
10^{27}$ years, with the neutrino mass $m_\nu \sim 10^{-6}\eV$ and an
arbitrary $R$. This possibility is quite reasonable and is in perfect
agreement with all experimental data and most theoretical predictions
for the present day. We have found that there is no point in discussing
$n > 4$ (see Fig. \ref{fig3}), at least in the context of
$0\nu2\beta$. If this happens to be the case, such a decay will be
practically forbidden.

One has to bear in mind, that all these results have meaning only in the
framework of the ADD model. In fact, as for today there is no
experimental hint, which supports such ideas. The only theoretical
motivation comes from string theory, but the realization of extra
dimensions may be of course completely different.

\section*{Acknowledgements}
This work is supported by grant no. 2P03B~071~25 from the Polish State
Committee for Scientific Research.

\end{document}